\documentclass[prl,showpacs,twocolumn,
amsmath,amssymb,superscriptaddress,floatfix,nofootinbib]{revtex4-1}
\usepackage[utf8]{inputenc}
\usepackage[sort&compress]{natbib}
\usepackage[normalem]{ulem}
\usepackage{lipsum} 
\usepackage{bm}
\usepackage{times}
\usepackage{amssymb,amsbsy,amsmath,amsfonts}
\usepackage{graphicx}
\usepackage{float}
\usepackage{color}
\usepackage{morefloats}
\usepackage{rotating}
\usepackage{srcltx}
\usepackage{slashed}
\usepackage{subfigure}
\usepackage{multirow}
\usepackage{verbatim}
\usepackage{hyperref}
\usepackage{tabularx}
\usepackage{adjustbox}
\usepackage[dvipsnames]{xcolor}
\usepackage{nicefrac}

\usepackage{hyperref}
\hypersetup{
	colorlinks	=true,
	urlcolor	=blue,
	linkcolor	=blue,
	citecolor	=blue,
	pdftitle	={},
	pdfauthor	={Jia-Ming Xie, Jun-Xu Lu, Li-Sheng Geng},
	pdfsubject	={Two-pole structures demystified: chiral dynamics at work}
	}

\begin{document}

\title{Two-pole structures as a universal phenomenon dictated by coupled-channel chiral dynamics}

\author{Jia-Ming Xie}
\affiliation{School of Physics, Beihang University, Beijing 102206, China}

\author{Jun-Xu Lu}
\email[Corresponding author: ]{ljxwohool@buaa.edu.cn}
\affiliation{School of Physics, Beihang University, Beijing 102206, China}

\author{Li-Sheng Geng}
\email[Corresponding author: ]{lisheng.geng@buaa.edu.cn}
\affiliation{School of
Physics,  Beihang University, Beijing 102206, China}
\affiliation{Peng Huanwu Collaborative Center for Research and Education, Beihang University, Beijing 100191, China}
\affiliation{Beijing Key Laboratory of Advanced Nuclear Materials and Physics, Beihang University, Beijing 102206, China }
\affiliation{Southern Center for Nuclear-Science Theory (SCNT), Institute of Modern Physics, Chinese Academy of Sciences, Huizhou 516000, Guangdong Province, China}

\author{Bing-Song Zou}
\affiliation{Peng Huanwu Collaborative Center for Research and Education, Beihang University, Beijing 100191, China}
\affiliation{CAS Key Laboratory of Theoretical Physics, Institute of Theoretical Physics,
Chinese Academy of Sciences, Beijing 100190, China}
\affiliation{School of Physical Sciences, University of Chinese Academy of Sciences, Beijing 100049, China}
\affiliation{School of Physics, Peking University, Beijing 100871, China}

\begin{abstract}
In the past two decades, one of the most puzzling phenomena discovered in hadron physics is that a nominal hadronic state can actually correspond to two poles on the complex energy plane. This phenomenon was first noticed for the $\Lambda(1405)$, then for $K_1(1270)$, and to a lesser extent for  $D_0^*(2300)$. In this Letter, we show explicitly how the two-pole structures emerge from the underlying universal chiral dynamics describing the coupled-channel interactions between heavy matter particles and pseudo Nambu-Goldstone bosons. In particular, the fact that two poles appear in between the two dominant coupled channels can be attributed to the particular form of the leading order chiral potentials of the  Weinberg-Tomozawa form.   Their lineshapes overlap with each other because the degeneracy of the two coupled channels is only broken by explicit chiral symmetry breaking of higher order.  We predict that for light-quark~(pion) masses heavier than their physical values (e.g., about 200 MeV in the $\Lambda(1405)$ case studied), the lower pole becomes a virtual state, which can be easily verified by future lattice QCD simulations. Furthermore, we anticipate similar two-pole structures in other systems, such as the isopin $1/2$
$\bar{K}\Sigma_c-\pi\Xi'_c$ coupled channel, which await for experimental discoveries. 
\end{abstract}


\maketitle

\textit{Introduction}.---Hadronic states that decay strongly are referred to as resonances. 
Experimentally, they often show up as enhancements in the invariant mass distributions of their decay products and are  parameterized with the Breit-Wigner~(BW) formalism. It is well known that close to two-body thresholds, however, the BW parameterization cannot accurately capture all the physics and may misrepresent the true dynamics~\cite{Gardner:2001gc,Mai:2022eur}. There are also other scenarios where more careful analyses are needed, one of which is where the enhancements may differ depending on the observing channels.  Among the latter, the $\Lambda(1405)$ state has attracted the most attention, and to a lesser extent, the $K_1(1270)$ and $D_0^*(2300)$ states~\cite{Meissner:2020khl}.

The $\Lambda(1405)$, with quantum numbers $J^P=1/2^-$, $I=0$ and $S=-1$, has remained puzzling in the constituent quark model~\cite{Capstick:1986ter} because  it is lighter than its nucleon-counterpart $N^*(1535)$ and  the mass difference between $\Lambda(1405)$ and its spin-partner $\Lambda(1520)$ with $J^P=3/2^-$ is much larger than the corresponding splitting in the nucleon sector~\cite{Hyodo:2011ur}. On the other hand, the $\Lambda(1405)$ was predicted to be a $\bar{K}N$ bound state even before its experimental discovery~\cite{Dalitz:1959dn}. Such a picture received further support in the chiral unitary approaches that combine SU(3)$_L$$\times$SU(3)$_R$  chiral dynamics and elastic unitarity (see Refs.~\cite{Oller:2000ma,Oller:2019opk} for a more complete list of references).  An unexpected finding of the chiral unitary approaches is that the $\Lambda(1405)$ actually corresponds to two dynamically generated poles on the second Riemann sheet of the complex energy plane~\cite{Oller:2000fj,Jido:2003cb}, between the thresholds of $\pi \Sigma(1330)$ and $\bar{K}N(1433)$, where the numbers in the brackets are the thresholds of the respective channels in units of MeV. Such a two-pole picture~\footnote{\label{footnote1}A proper definition of two-pole structures is needed for clarifying  this mysterious phenomenon. In this work, two-pole structures refer to the fact that two dynamically generated states, one resonant and one bound (with respect to the most strongly coupled channels), are located close to each other between two coupled channels  and have a mass difference smaller than the sum of their widths. As a result, the two poles overlap in the invariant mass distribution of their decay products, which creates the impression that there is only one state.} 
has recently been reconfirmed in the unified description of meson-baryon scattering at next-to-next-to-leading order~(NNLO)~\cite{Lu:2022hwm}. In the following years, it was shown that the  $K_1(1270)$~\cite{Roca:2005nm,Geng:2006yb} and $D_0^*(2300)$~\cite{Kolomeitsev:2003ac,Guo:2006fu,Guo:2009ct,Albaladejo:2016lbb,Guo:2018tjx,Du:2020pui}  also correspond to two poles~\footnote{In two recent works, it was shown that the $\Xi(1890)$~\cite{Molina:2023uko} and $b_1(1235)$~\cite{Clymton:2023txd} also correspond to two poles. The former is governed by the same chiral dynamics highlighted in the present work, while the latter is generated by a more complicated coupled-channel interaction.}, which are needed to explain many relevant experimental data~\cite{Geng:2006yb,Du:2020pui} or lattice QCD data~\cite{Asokan:2022usm}.

The fact that such two-pole structures emerge in three different sectors asks for an explanation. In  Ref.~\cite{Jido:2003cb}, it is shown that in the SU(3) flavor symmetry limit, one expects three bound states, one singlet and two degenerate octets.  In the physical world where SU(3) symmetry is broken, the singlet develops into the lower pole of the $\Lambda(1405)$, and one octet evolves into the higher pole.  Similar observations have been made for $K_1(1270)$~\cite{Roca:2005nm} and $D_0^*(2300)$~\cite{Albaladejo:2016lbb}. 

It is the purpose of the present Letter to explicitly demonstrate how the two-pole structures emerge from the underlying coupled-channel chiral dynamics and the pseudo Nambu-Goldstone~(pNG) nature of the pseudoscalar mesons. In particular, we would like to answer the following three questions. 1) whether the off-diagonal coupling between the two dominant channels plays a decisive role? 2) How explicit chiral symmetry breaking generates the two-pole structures? 3) Is the energy dependence of the Weinberg-Tomozawa~(WT) potential relevant?

\textit{Formalism}.---In this work, we focus on 
the two poles of $\Lambda(1405)$ and $K_1(1270)$ and highlight their common origins.~\footnote{We leave out the $D_0^*(2300)$ to the Supplemental Material.} We first spell out the leading order~(LO) chiral Lagrangians describing the  pseudoscalar-baryon~(PB) and pseudoscalar-vector~(PV) interactions~\footnote{As the two-pole structure persists up to higher chiral orders~\cite{{Ikeda:2012au,Guo:2012vv,Mai:2012dt,Lu:2022hwm,Zhou:2014ila}}, we stick to the
leading order to demonstrate the chiral dynamics at play and its universality in the present work.}, from which one can derive potentials $V$ of the WT type responsible for the dynamical generation of $\Lambda(1405)$ and $K_1(1270)$, and highlight their common feature. Then we briefly review the chiral unitary approaches.

For the PB interaction describing the scattering of a pseudoscalar meson off a ground-state octet baryon, the LO chiral Lagrangian  has the following form~\cite{Jido:2003cb}:
\begin{equation}
    \mathcal{L}_{PB}^{\mathrm{WT}}=\frac{1}{4f^2}\mathrm{Tr}\left(\Bar{\mathcal{B}}i\gamma^{\mu}\left[\Phi\partial_{\mu}\Phi-\partial_{\mu}\Phi\Phi,\mathcal{B}\right]\right),
\end{equation}
from which one can obtain the potential in the center of mass (c.m.) frame
\begin{equation}
    V_{ij}=-\frac{C_{ij}}{4f^2}\left(2\sqrt{s}-M_{i}-M_{j}\right)=-\frac{C_{ij}}{4f^2}\left(E_i+E_j\right),
    \label{VPB}
\end{equation}
where the subscripts $i$ and $j$ represent the incoming and outgoing  channels in isospin basis, $M$ is the mass of the baryon and $E$ is the energy of the pseudoscalar meson. $C_{ij}$ are the corresponding Clebsch-Gordan (CG) coefficients. Note that we have neglected the three momentum of the baryon in comparison with its mass and numerically verified that such an approximation has no impact on our discussion.

Likewise for the PV interaction, the LO chiral Lagrangian~\cite{Birse:1996hd,Roca:2005nm} is  
\begin{equation}
    \mathcal{L}_{PV}^{\mathrm{WT}}=-\frac{1}{4f^2}\mathrm{Tr}\left(\left[\mathcal{V}^{\mu},\partial^{\nu}\mathcal{V}_{\mu}\right]\left[\Phi,\partial_{\nu}\Phi\right]\right),
\end{equation}
from which one can  obtain the following potential projected onto $S$-wave,
\begin{equation}
\begin{aligned}
    V_{ij}\left(s\right)=&-\epsilon^i \cdot \epsilon^j \frac{C_{ij}}{8f^2}[ {3s-\left(M_i^2+m_i^2+M_j^2+m_j^2\right)}  \\ &  {-\frac{1}{s}\left(M_i^2-m_i^2\right)\left(M_j^2-m_j^2\right)} ].
    \label{VPV}
\end{aligned}
\end{equation}
Note that $M_{i,j}$  are masses of vector mesons and $m_{i,j}$ are those of pseudoscalar mesons. Close to threshold, considering the light masses of the pseudoscalar mesons as well as the chiral limit of $M_i=M_j\equiv M$, Eq.~(\ref{VPV}) can be simplified to
\begin{equation}
     V_{ij}\left(s\right)=-\epsilon^i \cdot \epsilon^j \frac{C_{ij}}{8f^2}4M\left(E_i+E_j\right),
\end{equation}
which is  the same as Eq.~(\ref{VPB}) up to the scalar product of polarization vectors, trivial dimensional factors, and CG coefficients.

In the chiral unitary approaches~\cite{Oller:2000ma},  the unitarized amplitude reads
 \begin{equation}
    T=\left(1-VG\right)^{-1}V,
\end{equation}
 where $G$ is a diagonal matrix with elements $G_{kk}\equiv G_{k}\left(\sqrt{s}\right)$.  The loop function $G_k\left(\sqrt{s}\right)$ of channel $k$ is logarithmically divergent and can be regulated either in the dimensional regularization (DR) scheme  or the cutoff scheme. In the DR scheme, a subtraction constant is introduced, while in the cutoff scheme one needs a cutoff. In practice, the subtraction constants or cutoff values are determined by fitting to the scattering data but should be of natural size in order for the chiral unitary approaches to make sense. For details, see, e.g., Refs.~\cite{Oller:2000ma,Roca:2005nm}.



The couplings of a resonance/bound state to its constituents can be obtained from the residues of the corresponding pole on the complex energy plane, \textit{i.e.},
\begin{equation}
    g_i g_j=\lim_{\sqrt{s} \rightarrow z_R}\left(\sqrt{s}-z_R\right)T_{ij}\left(\sqrt{s}\right),
    \label{coupling}
\end{equation}
where $z_R\equiv m_R-i\Gamma_R/2$ is the pole position.

\begin{figure}[htpb]
    \includegraphics[width=3.4in]{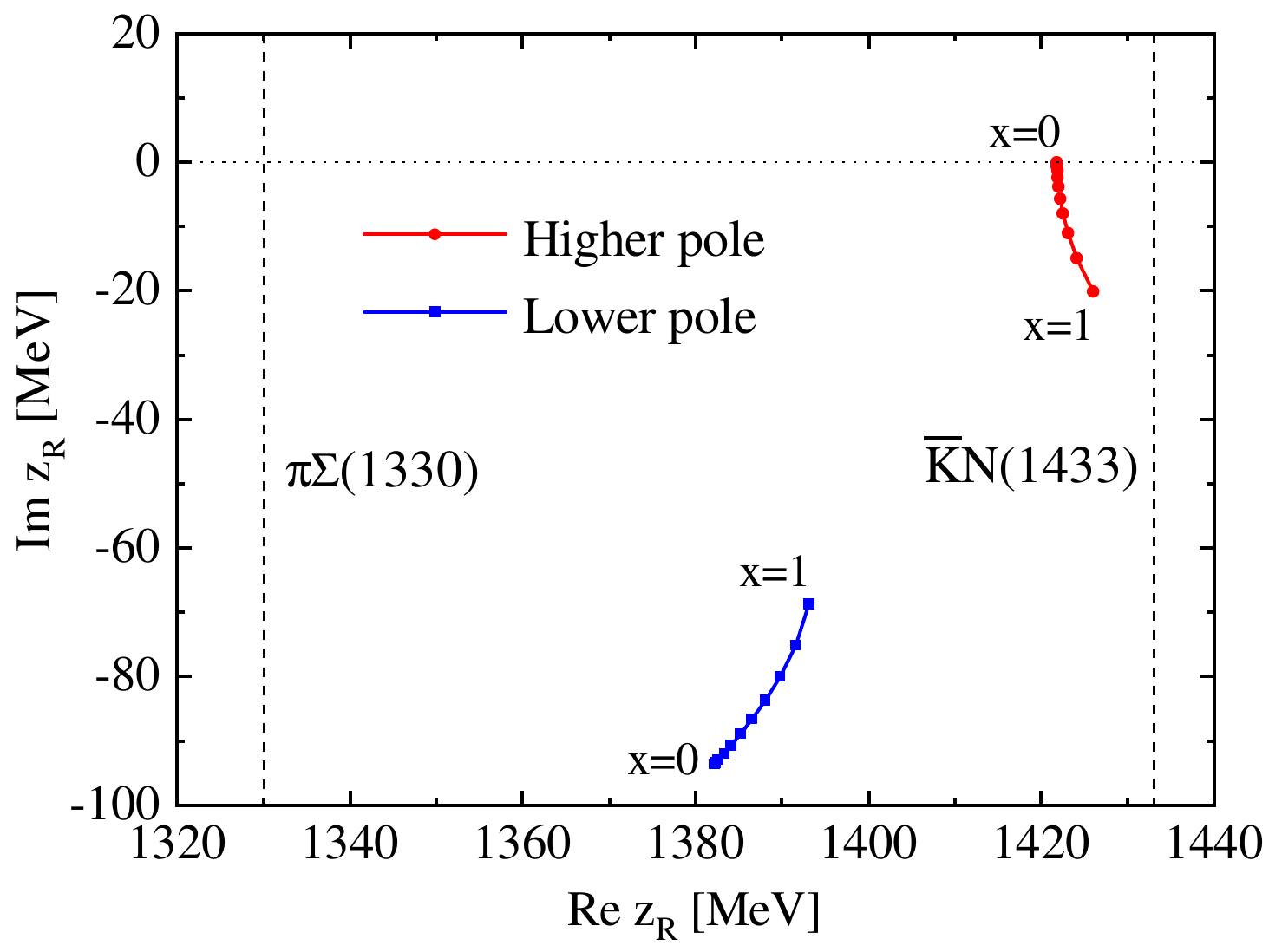}
    \caption{Evolution of the two poles of $\Lambda\left(1405\right)$ as a function of the off-diagonal potential $x \times V_{\bar{K}N-\pi\Sigma}$ with $0\le x\le 1$. Every point on the lines is taken in steps of $x=0.1$.}
    \label{fig:trajectory}
\end{figure}
\textit{Coupled channel effects}.---We first focus on the $\Lambda(1405)$ state. In the isospin $0$ and  strangeness $-1$ meson-baryon system, the $\bar{K}N$ and $\pi\Sigma$ channels play the most important role around the 1400 MeV region~\cite{Jido:2003cb,Hyodo:2007jq}. With the following subtraction constants $a_{\bar{K}N}=-1.95$ and $a_{\pi \Sigma}=-1.92$,
we find  two poles on the complex energy plane, \textit{i.e.}, $W_H=1426.0-20.1i$ MeV and $W_L=1393.1-68.7i$ MeV, consistent with the LO~\cite{Jido:2003cb}, next-to-leading order~(NLO)~\cite{Ikeda:2012au,Guo:2012vv,Mai:2012dt}, and NNLO results~\cite{Lu:2022hwm}.
\begin{figure*}[htpb]
    \centering
    \includegraphics[width=3.2in]{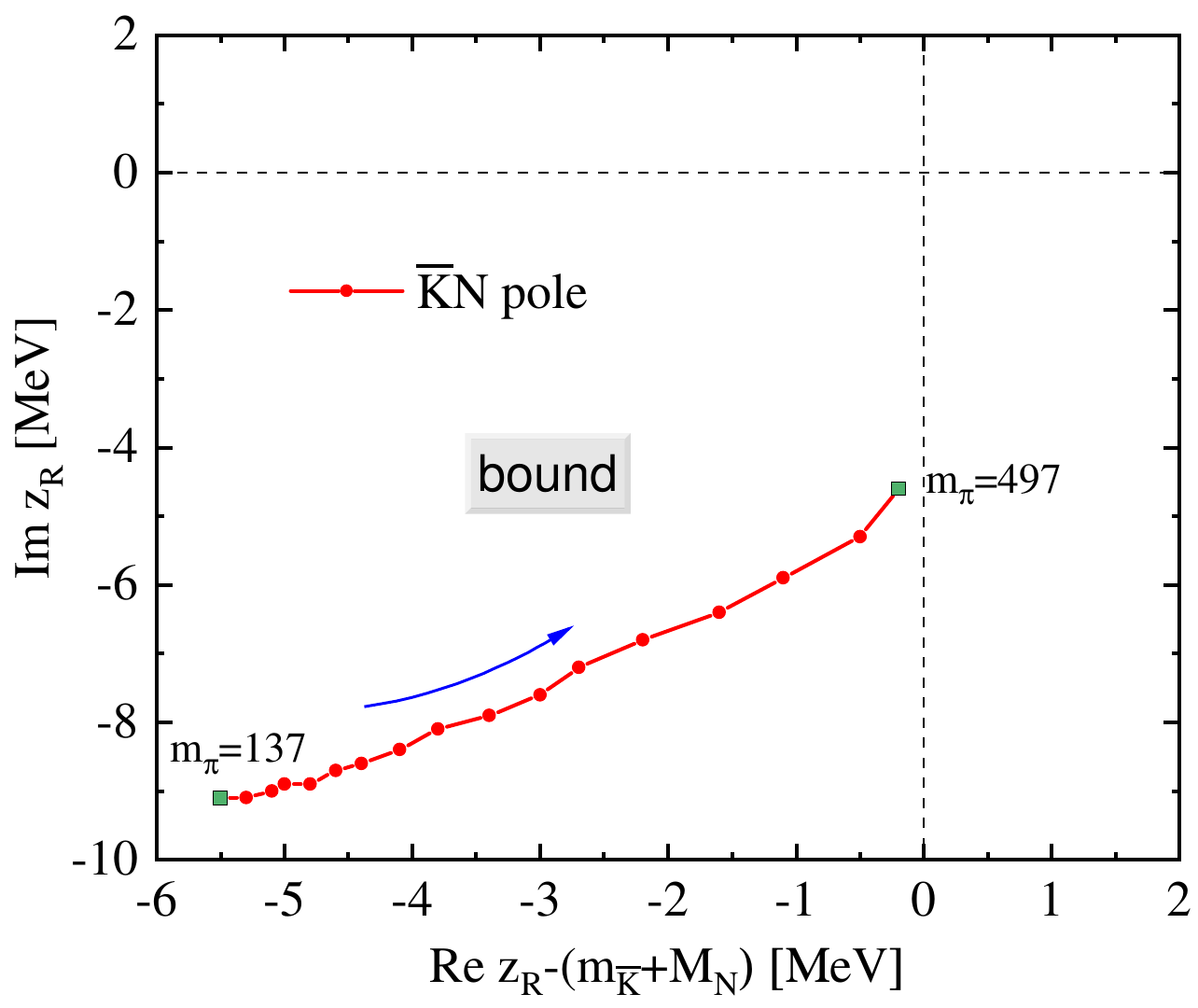}\quad \quad \quad
       \includegraphics[width=3.3in]{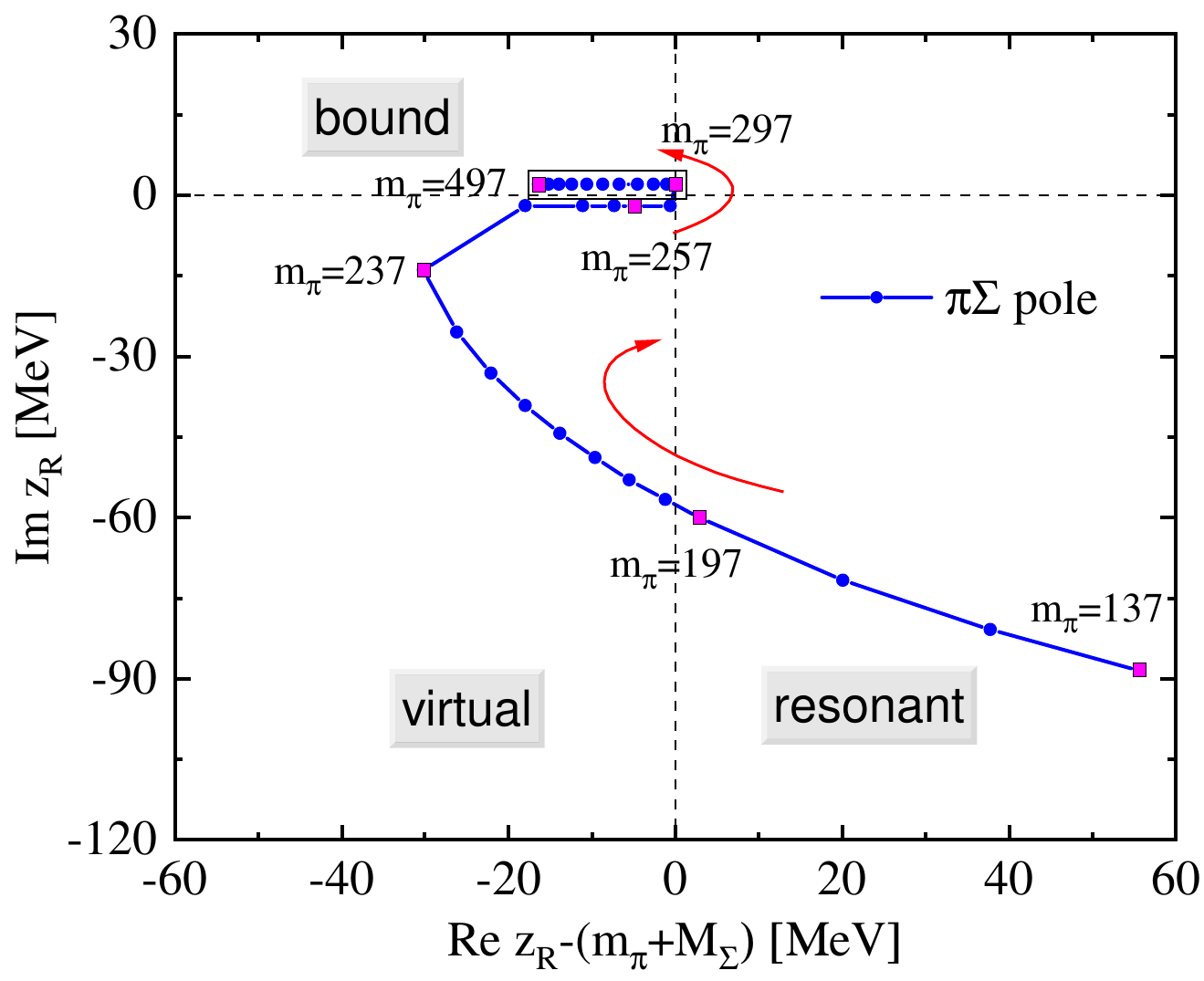}
    \caption{Trajectories of the  two poles of $\Lambda(1405)$ as functions of the pion mass  $m_{\pi}$ from 137 MeV to 497 MeV. Critical masses  are labeled by solid squares, between which  the points are equally spaced.}
    \label{fig:2channel}
\end{figure*}
One might naively expect that the two poles are linked to the coupling between the $\bar{K}N$ and $\pi\Sigma$ channels. 
This is actually not the case. To demonstrate this, we decrease the coupling between $\Bar{K}N$ and $\pi \Sigma$ by multiplying a factor $0\le x\le1$ to the  off-diagonal matrix elements of the WT potential and obtain the evolution of the two poles  shown in Fig.~\ref{fig:trajectory}.
Two things are noteworthy. First, even in the limit
of complete decoupling, \textit{i.e.}, $x = 0$, the two poles still appear in between
the $\Bar{K}N$ and $\pi\Sigma$ thresholds, but the imaginary part of the higher pole approaches zero while the imaginary part of the lower pole becomes larger. 
 Second, the coupling between the two channels not only pushes the two poles higher, but also allows the higher pole to decay into the $\pi \Sigma$ channel and as a result develops a finite width. Nevertheless the most important issue to note is that the coupling between the two channels is not the driving factor for the existence of two dynamically generated states in between the two relevant channels. On the other hand, it does play a role in the development of the two-pole structure, because otherwise the higher pole will not manifest itself in the invariant mass distribution of the lower $\pi\Sigma$ channel. We note that  Ref.~\cite{Cieply:2016jby} has used a similar approach, the so-called zero coupling limit~\cite{Hyodo:2007jq}, to study the pole contents of various unitarized chiral approaches.

\textit{Explicit chiral symmetry breaking}.---In the following, we explicitly show that it is the underlying chiral dynamics that is responsible for the emergence of the two-pole structure.   According to Eq.~(\ref{VPB}), the diagonal WT interaction  is proportional to the energy of the pseudoscalar meson. For the $\Bar{K}N$ and $\pi\Sigma$ channels of our interest, they read~\footnote{As we have shown that the coupling between $\bar{K}N$ and $\pi\Sigma$ does not play a decisive role in generating the two-pole structure, the following discussion should be understood in the single-channel approximation.} 
\begin{equation}
\begin{aligned}
    V_{\Bar{K}N- \Bar{K}N}\left(\sqrt{s}\right)=&-\frac{6}{4f^2}E_{\Bar{K}}=-\frac{6}{4f^2}\sqrt{m_{\Bar{K}}^2+q_{\Bar{K}}^2},\\
    V_{\pi \Sigma- \pi \Sigma}\left(\sqrt{s}\right)=&-\frac{8}{4f^2}E_{\pi}=-\frac{8}{4f^2}\sqrt{m_\pi^2+q_{\pi}^2}.
    \label{V-diagonal}
\end{aligned}
\end{equation}
Due to the explicit chiral symmetry breaking,  the mass of the kaon is much larger than that of the pion. As a result, close to threshold, the $\bar{K}N$ interaction is stronger than the $\pi \Sigma$ one, which leads to a $\bar{K}N$ bound state. In addition, the energy dependence and the small pion mass together enhance the $q^2$ term of the $\pi\Sigma$ interaction and therefore are responsible for the existence of a $\pi\Sigma$ resonance.~\footnote{The different role played by  the heavy kaon and the light pion has been noted previously in other contexts, see, e.g., Refs.~\cite{MartinezTorres:2011gjk,Jido:2011tb}.} We stress that the role of explicit symmetry breaking can be appreciated by studying  the pole trajectories as a function of the light-quark (pion) mass.

As the pion mass changes, masses of the baryons and the kaon also vary. We adopt  the covariant baryon chiral perturbation theory to describe their light-quark mass dependence. Up to  $\mathcal{O}\left(p^2\right)$, the octet baryon masses read
\begin{equation}M_B\left(m_{\pi}\right)=M_0+M_B^{\left(2\right)}=M_0+\sum_{\phi=\pi, K}\xi_{B, \phi}m_{\phi}^2,
\end{equation}
where $M_0$ is the chiral limit baryon mass and $\xi_{B, \phi}$ are the relevant coefficients that contain three low-energy constants, which are fitted to the lattice QCD data of the PACS-CS Collaboration~\cite{PACS-CS:2008bkb} in Ref.~\cite{Ren:2012aj}, where one can also find the pion mass dependence of the kaon.  

The trajectories of the two poles of $\Lambda(1405)$ are shown in Fig.~\ref{fig:2channel}. The evolution of the higher pole is simple. As the pion mass increases, both its real and imaginary parts  decrease. This indicates that the effective $\bar{K}N$ interaction and coupling to $\pi\Sigma$ both decrease as the pion (kaon) mass increases. Note that as the pion mass increases, the two thresholds increase as well. On the other hand, the trajectory of the lower pole is more complicated and highly nontrivial. As the pion mass increases, it first becomes a virtual state from a resonant state for a pion mass of about 200 MeV. For a pion mass of about 300 MeV, it becomes a bound state and remains so up to the pion mass of 500 MeV. The evolution of the lower pole  clearly demonstrates the chiral dynamics underlying the two-pole structure of $\Lambda(1405)$.

To check how the energy dependence of the chiral potential affects the two-pole structure~\footnote{The relation between the energy dependence of the WT potential and the simultaneous appearance of a bound state and a resonant state has been noted in, e.g., Ref.~\cite{Ikeda:2010tk}.}, we  replace $E_i+E_j$ of the chiral potential of Eq.~(\ref{VPB}) with $m_i+m_j$. With the original subtraction constants, we obtain only one pole at $1413.3-13.2i$ MeV, corresponding to a $\bar{K}N$ bound state. 
 We checked that switching off the off-diagonal interaction affects little our conclusion. As the pion mass is much smaller than kaon mass, the attraction of the $\pi \Sigma$ single channel is weaker than that of the $\Bar{K}N$ single channel, thus cannot support a bound state. Of course if we increase the strength of the attractive potential, we can obtain two bound states, but not a bound state and a resonant state, and as a result, there is no two-pole structure any longer.

 $K_1(1270)$.---From the above study, one immediately realizes that if one replaces the matter particles (the ground-state baryons) with the ground-state vector mesons, one may also expect the existence of a two-pole structure. This is indeed the case as shown in Refs.~\cite{Roca:2005nm,Geng:2006yb}, where $K_1(1270)$ is found to correspond to two poles. The most relevant channels are  $K^* \pi(1030)$ and  $\rho K(1271)$.
In Ref.~\cite{Geng:2006yb}, it was shown that with $\mu=900$ MeV,   $a\left(\mu\right)=-1.85$, and $f=115$ MeV,
where $\mu$ is the renormalization scale, $a\left(\mu\right)$ is the common subtraction constant, and $f$ is the  pion decay constant, one finds two poles located at  $W_H=1269.3-1.9i$ MeV and  $W_L=1198.1-125.2i$ MeV below the $\rho K$ and above the $K^* \pi$ thresholds.  Eliminating  the three  higher channels, we can find almost the same two poles located at  $W_H=1269.5-12.0i$ MeV and $W_L=1198.5-123.2i$ MeV,
by adjusting slightly the  subtraction constants as $a_{K^{\ast}\pi}=-2.21$, $a_{\rho K}=-2.44$.

\begin{figure}[htpb]
    \includegraphics[width=3.4in]{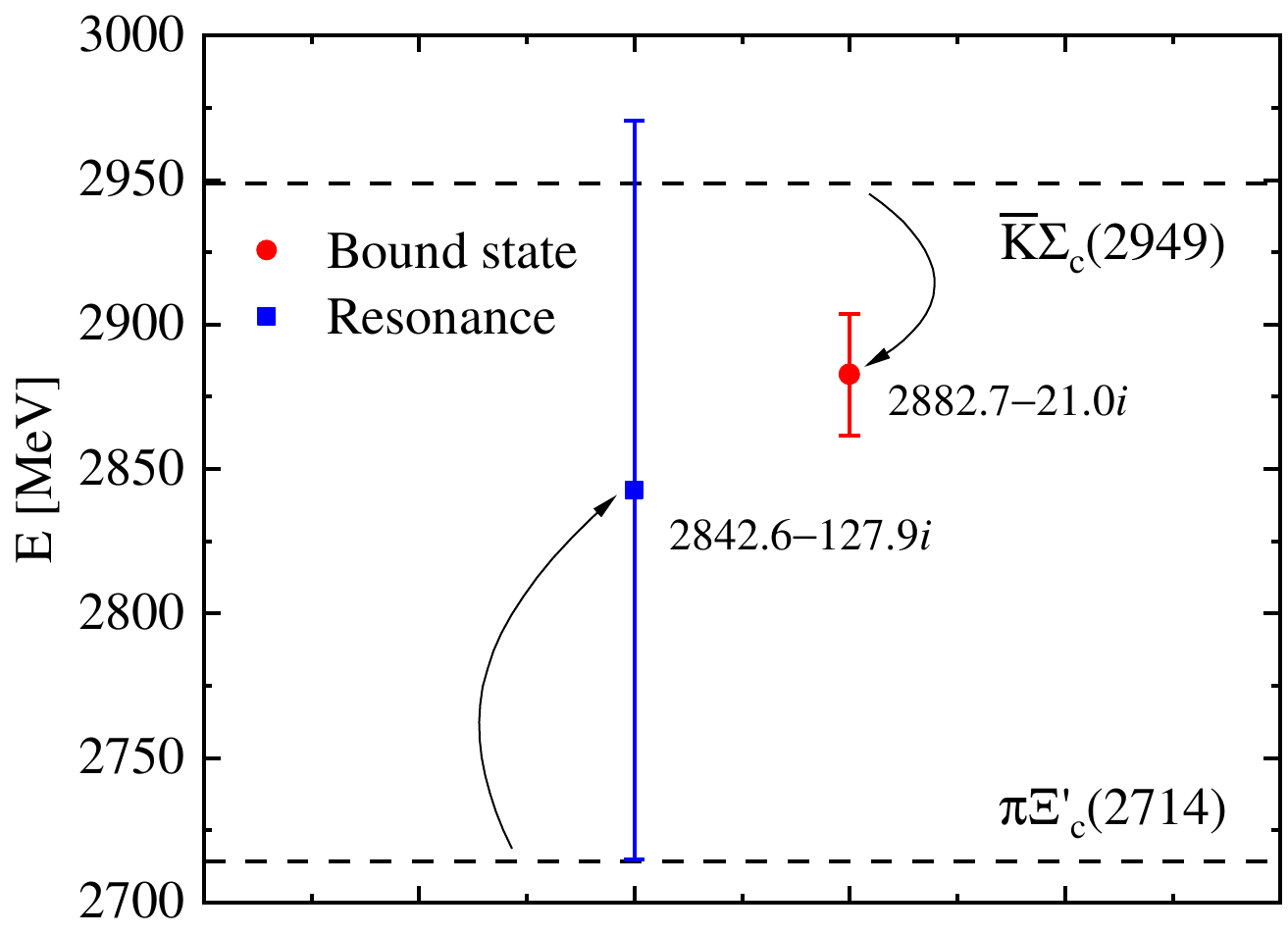}
    \caption{Two poles of the $\Bar{K}\Sigma_c-\pi \Xi_c'$ system. 
    The dominant channels in relation to the two states are denoted by the arrows. The vertical bars are the widths corresponding to twice of the imaginary parts of the pole positions.}
    \label{fig:single charm baryon}
\end{figure}
\begin{figure}[htpb]
    \includegraphics[width=3.4in]{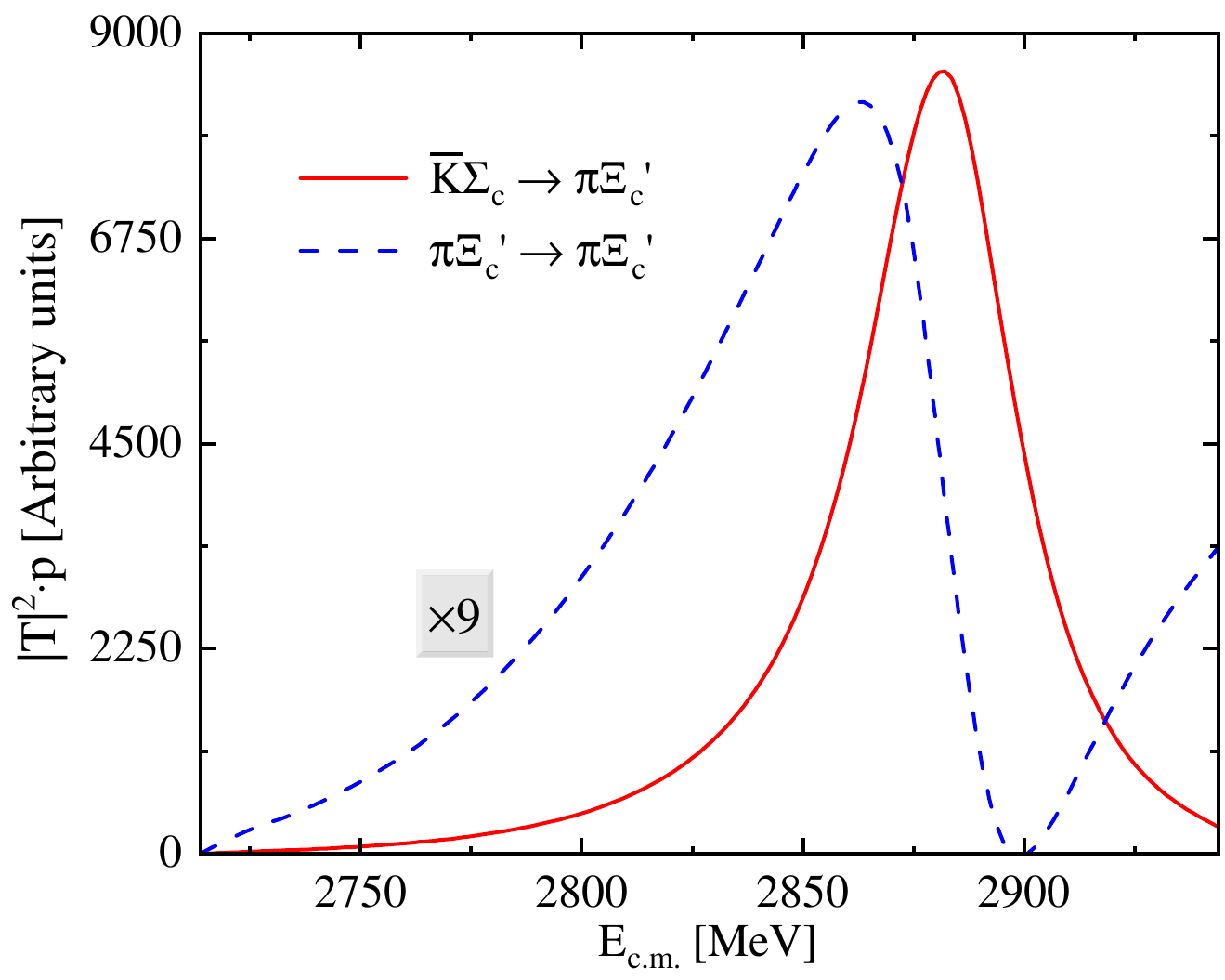}
    \caption{Invariant mass distributions of $\pi \Xi_c'$  (in arbitrary units) as functions of the c.m. energy,  $\left|T_{\Bar{K}\Sigma_c \rightarrow \pi\Xi_c'}\right|^2 p_{\pi}$ (red solid) and $9\times \left|T_{\pi\Xi_c' \rightarrow \pi\Xi_c'}\right|^2 p_{\pi}$ (blue dashed), where $p_{\pi}$ is  the 3-momentum of the pion in the c.m. frame of the final states.}
    \label{fig:double peak}
\end{figure}
\textit{Further two-pole structures.}---
In principle, because of the universality of chiral dynamics discussed in this work, one can expect more such two-pole structures in other systems composed of a pair of heavy matter particles and pseudoscalar mesons, such as the singly charmed baryon sector~\cite{Lu:2014ina}. Using the criteria proposed in this work, one can identify the two channels $\Bar{K}\Sigma_c(2949)$ and $\pi\Xi_c'(2714)$ that has the potential of generating a  two-pole structure. 

With a common cutoff of $\Lambda=800$ MeV in the cutoff regularization scheme and using the WT potential similar to Eq.~(\ref{VPB})~\footnote{It is hard to directly compare with Ref.~\cite{Yu:2018yxl}, because there  the extra coupled channels of charmed mesons and ground-state baryons are considered,  which are not constrained by the same chiral dynamics studied in this work.}, we find two poles in the isospin $1/2$ channel, located at $W_H=2882.7-21.0i$ MeV and $W_L=2842.6-127.9i$ MeV. The higher pole couples strongly to the $\Bar{K}\Sigma_c$  channel, while the lower pole couples more to the $\pi\Xi_c'$ channel, as shown in Fig.~\ref{fig:single charm baryon}. 
With the unitary amplitudes  $\Bar{K}\Sigma_c \rightarrow \pi\Xi_c'$ and $\pi\Xi_c' \rightarrow \pi\Xi_c'$ and following Ref.~\cite{Lu:2014ina}, we can construct the $\pi \Xi_c'$ invariant mass distributions shown in Fig.~\ref{fig:double peak}. Note that the lineshape of the $\Bar{K}\Sigma_c \rightarrow \pi\Xi_c'$ amplitude overlaps much with that of the $\pi\Xi_c' \rightarrow \pi\Xi_c'$ one, but they peak at slightly different positions and have different widths. These can be qualitatively explained by the fact that the  process $\Bar{K}\Sigma_c \rightarrow \pi\Xi_c'$ receives more contribution from the higher pole, while the process $\pi\Xi_c' \rightarrow \pi\Xi_c'$ couples more to the lower pole. We need to stress that although the two-pole structure is tied to the underlying chiral dynamics, the regularization, \textit{i.e.}, the cutoff in the present case, plays a relevant role. We therefore encourage further theoretical and experimental studies of the states predicted.

\textit{Conclusion and outlook}.---  We have examined  the origin of the mysterious two-pole structures and attributed this fascinating phenomenon to the underlying  chiral dynamics. First, chiral symmetry strongly constrains the interactions of a matter particle with a pseudoscalar meson, which are often referred to as the Weinberg-Tomozawa potentials. Second, the pseudo Nambu-Goldstone boson nature of $\pi$, $K$, and $\eta$  are responsible for the generation of two nearby poles: one bound and one resonant. Furthermore, the explicit chiral  and SU(3) flavor symmetry breaking dictates that the two relevant coupled channels are close to each other such that the lineshapes of the two states overlap and thus create the impression that there is only one state.  We anticipate more such two-pole structures in other systems governed by the same chiral dynamics and encourage dedicated experimental and lattice QCD studies to verify the chiral dynamics underlying such phenomena.  Last, we stress that flavor symmetry also plays a relevant role here, as it dictates the relative coupling strengths between different channels. 



\textit{Acknowledgment}.---L.S.G. thanks Qiang Zhao, Ulf Mei{\ss}ner, Eulogio Oset, Antonio Jose Oller, Daisuke Jido, and Tetsuo Hyodo for enlightening communications. J.M.X. is grateful to Fang-Zheng Peng for useful discussions. This work is partly supported by the National Natural Science Foundation of China under Grant  No.11975041 and No.11961141004. J.X.L. acknowledges support from the National Natural Science Foundation of China under Grant No.12105006. B.S.Z was supported by the National Natural Science Foundation of China (NSFC) and the Deutsche Forschungsgemeinschaft (DFG, German Research Foundation) through the funds provided to the Sino-German Collaborative Research Center TRR110 Symmetries and the Emergence of Structure in QCD (NSFC Grant No. 12070131001, and DFG Project-ID 196253076-TRR 110), the NSFC (11835015, and 12047503), and the Grant of Chinese Academy of Sciences (XDB34030000).

\bibliography{bib.bib}

\clearpage
\begin{widetext}
\setcounter{page}{1}
\section{Supplemental material}
In this Supplemental Material, we provide more examples of two-pole structures, which, although are not as compelling or clear cut as  the three cases detailed in the main text, are more or less driven by the same chiral dynamics. One is $D_0^*(2300)$ and the other are dynamically generated by the meson-baryon coupled-channel interactions  of strangeness $-2$ and 0. In addition, we comment on $f_0(500)$ and $f_0(980)$ and how they resemble but do not belong to two-pole structures studied in the main text. Finally we discuss the new  study of the Baryon Scattering Collaboration compared with our results.

\subsection{$D_0^*(2300)$}
Following the studies on $\Lambda(1405)$ and $K_1(1270)$, one can think of replacing the ground-state baryons with the charmed mesons $D$ and $D_s$ and anticipate the appearance of two-pole structures as well. This is   indeed the case, see, e.g., Ref.~\cite{Albaladejo:2016lbb}.  However, there is a complicating factor. That is to say, unlike the three cases discussed in the main text, because of the closeness of the $D\eta(2415)$ and $D_s\bar{K}(2464)$ channels, now three channels are at play, \textit{i.e.}, $D\pi(2005)$, $D\eta(2415)$, and $D_s\bar{K}(2464)$. Nonetheless, the same chiral dynamics works and dynamically generates two poles.
 The LO chiral Lagrangian responsible for the coupled channel interaction between a charmed meson and a pNG boson~\cite{Guo:2006fu} is
\begin{equation}
    \mathcal{L}_{PP}^{\mathrm{WT}}=\frac{1}{4f^2}\left(\partial^{\mu}\mathcal{P}\left[\Phi,\partial_{\mu}\Phi\right]\mathcal{P}^{\dagger}-\mathcal{P}\left[\Phi,\partial_{\mu}\Phi\right]\partial^{\mu}\mathcal{P}^{\dagger}\right),
    \label{LPP}
\end{equation}
from which one can obtain the same $S$-wave potential as Eq.~(\ref{VPV}), 
except for the absence of the product of polarization vectors of vector mesons.
Note that $M_{i,j}$  are the masses of  $D$ and $D_s$ mesons and  $m_{i,j}$ are those of pNG bosons. With $\mu=m_D=1867$ MeV, $a_{D\pi}=a_{D\eta}=a_{D_{s}\bar{K}}=-0.63$, we find a  lower pole  located at $W_L=2100.0-100.9i$ MeV and a higher pole located at $W_H=2439.8-43.1i$  MeV between the $D\eta$ and $D_{s}\bar{K}$ thresholds. We stress that in the present case, the $D\eta$ channel plays a relevant role as well because of its closeness to the $D_s\bar{K}$ channel and its large couplings to the two poles. 

Another interesting observation is that with the LO contribution~\cite{Guo:2006fu}, the width of the higher pole is much smaller than that obtained with the NLO contribution~\cite{Albaladejo:2016lbb}. Therefore, at LO, the two poles can actually be distinguished from each other in the invariant mass distribution of the $D\pi$ channel, thus making this two-pole structure different from those of the $\Lambda(1405)$ and $K_1(1270)$. 


\subsection{Other possible two-pole structures related to $\Lambda(1405)$  by SU(3) flavor symmetry}
 In the $S=-2$ meson-baryon system, with four channels $\pi \Xi(1455)$, $\Bar{K}\Lambda(1610)$, $\Bar{K}\Sigma(1688)$ and $\eta \Xi(1865)$, and a common cutoff $\Lambda=630$ MeV of natural size, two poles can be dynamically generated: $W_L=1567.8-127.4i$ MeV and $W_H=1687.0-0.8i$ MeV. Although at first sight it seems that they can be identified as  the $\Xi(1620)$ and $\Xi(1690)$ states~\cite{ParticleDataGroup:2022pth}, both the width and the mass of the low-energy pole are quite different from those of $\Xi(1620)$ and $\Xi(1690)$, making such an assignment questionable. We note that the LO study of Ref.~\cite{Ramos:2002xh} predicts a state corresponding to the $\Xi(1620)$ but a cusp structure relevant to the $\Xi(1690)$. In Ref.~\cite{Garcia-Recio:2003ejq}, similar to our present study, two states are dynamically generated with the LO WT potential, while the predicted width of the lower pole is much larger than the width of $\Xi(1620)$ and ours. In Ref.~\cite{Feijoo:2023wua}, including the Born terms and NLO contributions, two states similar to ours are found but the width of the higher-pole is closer to the experimental width of $\Xi(1690)$. We conclude that although two poles can be dynamically generated in the chiral unitary approaches, the assignments of them to $\Xi(1620)$ and $\Xi(1690)$ are premature.
 
 In the $S=0$ meson-baryon sector, with four channels $\pi N(1075)$, $\eta N(1485)$, $K\Lambda(1610)$ and $K\Sigma(1688)$, and a common cutoff $\Lambda=650$ MeV of natural size,  one can dynamically generate two poles with only the WT interaction. The higher pole is located at $W_H=1506.8-80.2i$ MeV, which could be identified as the $N^*(1535)$ state. The lower pole $W_L=1162.3-159.5i$ MeV is located above the $\pi N$ threshold. We note that our results are qualitatively similar to those of Ref.~\cite{Chen:2022zgm}.

\subsection{$f_0(500)$ and $f_0(980)$}
As a matter of fact, $f_0(500)$ and $f_0(980)$ are widely accepted as $\pi\pi$ and $K\bar{K}$ molecules. From this perspective, they resemble the two poles of $\Lambda(1405)$ or $K_1(1270)$. On the other hand, because the mass thresholds of $\pi\pi$ and $K\bar{K}$ are well separated, $f_0(500)$ and $f_0(980)$ can be easily recognized as two different states in the $\pi\pi$ invariant mass distributions. As a result, we do not classify them as a two-pole structure. 
\section{Light-quark mass evolution of the two poles of $\Lambda(1405)$ following a different trajectory}
\begin{figure*}[htpb]
    \centering
    \includegraphics[width=3.2in]{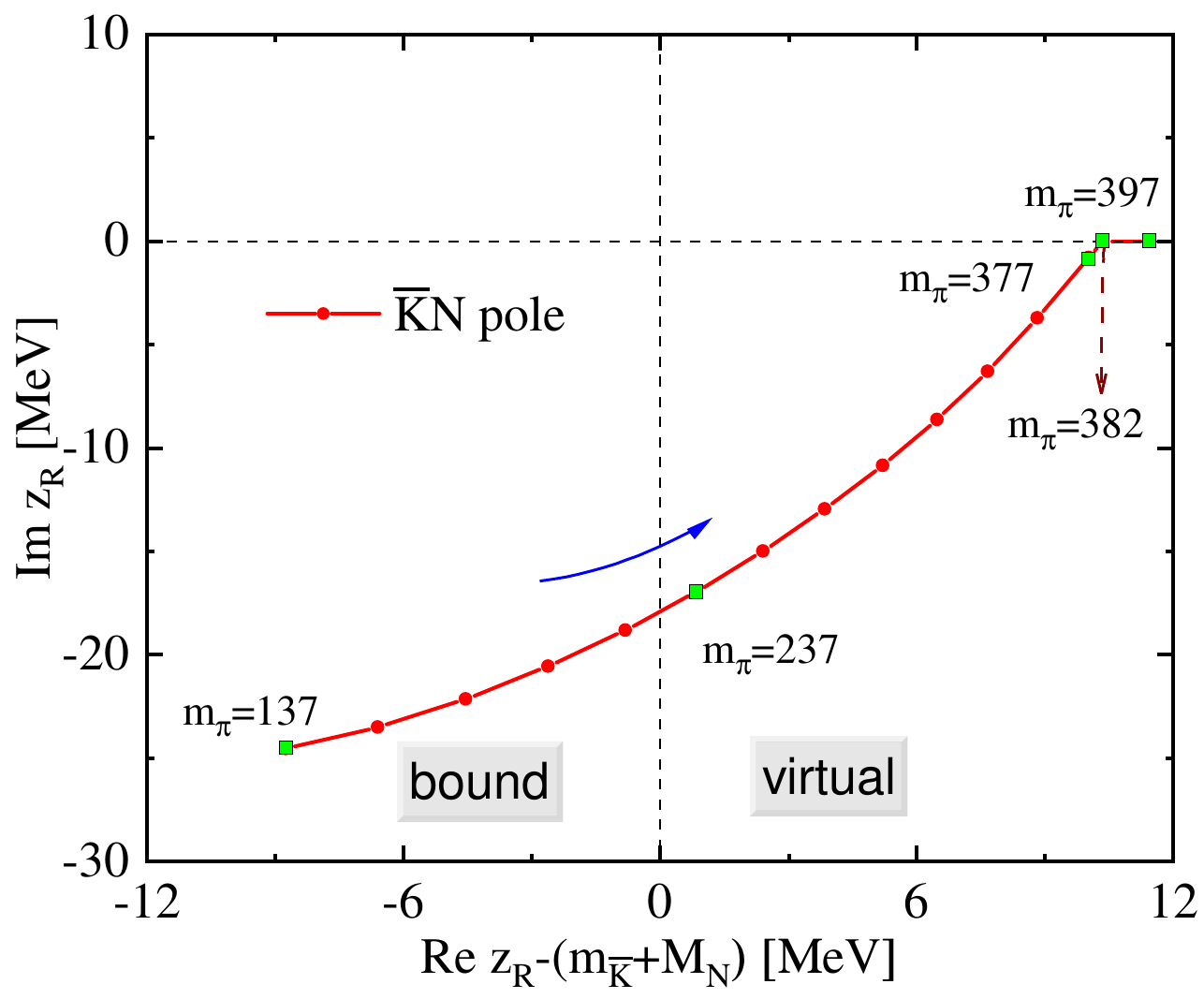}\quad
       \includegraphics[width=3.27in]{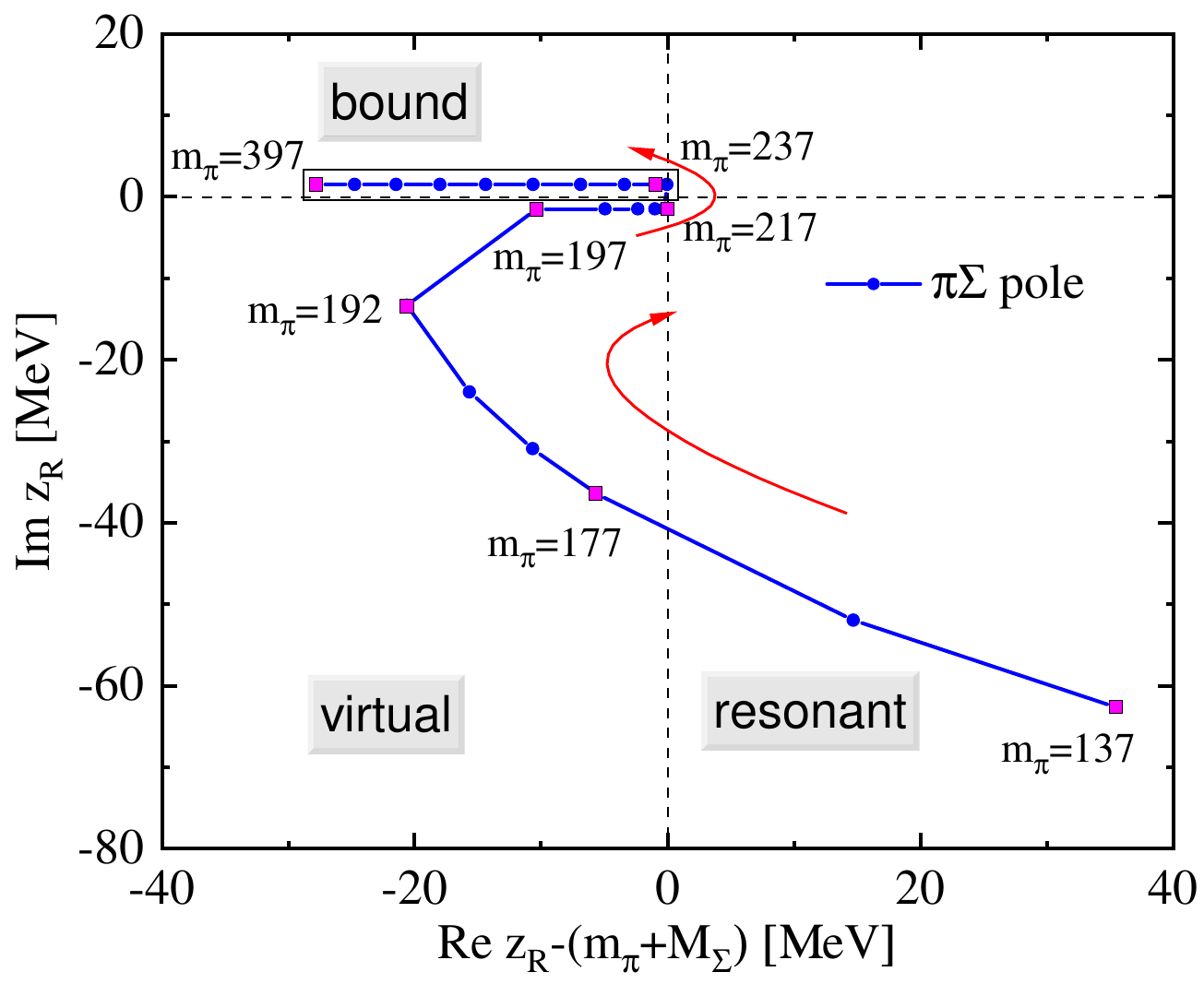}
    \caption{Trajectories of the  two poles of $\Lambda(1405)$ as functions of the pion mass  $m_{\pi}$ from 137 MeV to 397 MeV following the QCDSF-UKQCD light-quark mass evolution. Critical masses  are labeled by solid squares, between which  the points are equally spaced.}
    \label{fig:2channel2}
\end{figure*}

According to  Fig.~2 of the main text, the pion mass at which the lower pole evolves from a resonant state to a  virtual \textit{resonant} state occurs at $m_{\pi}\approx$ 200 MeV and $m_K\approx$ 546 MeV. On the other hand, the latest lattice QCD study by the Baryon Scattering Collaboration (BaSC)~\cite{BaryonScatteringBaSc:2023ori,BaryonScatteringBaSc:2023zvt}  found two poles located at $E_1=1392(9)(2)(16)$ MeV and $E_2=1455(13)(2)(17)-i11.5(4.4)(4)(0.1)$ MeV, corresponding to a virtual \textit{bound} state (with respect to $\pi \Sigma$ and a resonant state (with respect to $\pi\Sigma$ or a bound state with respect to $\bar{K}N$), respectively. 
However, one must note that  the trajectories shown in Fig.~2 of the main text rely on the meson and baryon masses as a function of the masses of light quarks $u$, $d$, and $s$, or of the light mesons $\pi$ and $K$.  They are determined by fitting to the lattice QCD data of the PACS-CS Collaboration~\cite{PACS-CS:2008bkb,Ren:2012aj,Song:2018qqm}with the NLO baryon chiral perturbation theory. The recent lattice QCD study~\cite{BaryonScatteringBaSc:2023ori,BaryonScatteringBaSc:2023zvt}, on the other hand, was performed at $m_{\pi}\approx$ 200 MeV and $m_K\approx$ 487 MeV with the CLS D200 configuration.  As a result, it is not surprising that they do not coincide with each other.  Without a detailed knowledge on the light-quark mass dependence of the hadron masses and other physical quantities involved, such as the decay constants and the subtraction constants, in principle one cannot predict the trajectories which one particular lattice QCD simulation will follow.

For the results of the BaSC, one can reproduce their results obtained with the CLS D200 configuration with slightly tuned subtraction constants $a_{\bar{K}N}=-2.2$ and $a_{\pi\Sigma}=-1.7$ and all the other unspecified low-energy constants unchanged. Two poles emerge at $E_1'=1395.6$ MeV and $E_2'=1464.0-i24.5$ MeV, in reasonable agreement with the BaSC results. We note that the CLS D200 configuration fall almost on the QCDSF-UKQCD trajectories~\cite{Bietenholz:2011qq}, see Table~\ref{tab:d200}. We can fit the QCDSF-UKQCD results with the NNLO baryon chiral perturbation theory. At this order, the octet baryon masses~\cite{MartinCamalich:2010fp} read
\begin{equation}
M_B(m_{\pi})=M_0+M_B^{(2)}+M_B^{(3)}=M_0+\sum_{\phi=\pi,K}\xi^{(2)}_{B,\phi}m_{\phi}^2+\frac{1}{\left(4\pi F_{\phi}\right)^2}\sum_{\phi=\pi,K,\eta}\xi^{(3)}_{B,\phi}H_{B}(m_{\phi}).
\end{equation}
By a least-squares fit, we obtain $m_K^2=0.252658-0.489594m_{\pi}^2$~(in units of GeV)~\cite{Ren:2012aj} and the following LECs: $M_0=1022.8$ MeV, $b_0=2.0047$ GeV$^{-1}$, $b_D=0.060083$ GeV$^{-1}$, and $b_F=-0.070261$ GeV$^{-1}$. The achieved  $\chi^2/$d.o.f is about 1 for  28~(=32-4) degrees of freedom. With this light-quark mass dependence, we show  the trajectories of the two poles of $\Lambda(1405)$ as functions of the pion mass $m_{\pi}$ in Fig.~\ref{fig:2channel2}. One can see that at $m_{\pi}$=197 MeV the width of the lower pole vanishes and thus it becomes a virtual \textit{bound} state, in better agreement with the BaSC result~\cite{BaryonScatteringBaSc:2023ori,BaryonScatteringBaSc:2023zvt}.

Special cautions should be taken in interpreting the trajectories shown in either Fig.~2 of the main text or Fig.~1 of this Supplemental Material. In particular, the trajectories should be viewed only qualitatively because they are determined by the assumed light-quark mass dependence of the involved physical quantities. In particular, as a one-dimensional plot, the kaon mass dependence on the pion mass is nontrivial, which is quite different between the PACS-CS  and the QCDSF-UKQCD simulations.  In addition, in both plots, we have neglected the light-quark mass dependence of the decay constants and subtraction constants, which may not be totally negligible, particularly the latter. 

\begin{table*}[!ht]
    \caption{Hadron masses predicted by the NNLO ChPT fit to the QCDSF-UKQCD data~\cite{Bietenholz:2011qq} in comparison with the CLS D200 data~\cite{BaryonScatteringBaSc:2023ori,BaryonScatteringBaSc:2023zvt} at $m_{\pi}\approx 200$~(in units of MeV).\label{tab:d200}}
    \begin{tabular}{c|cccc}
\hline\hline
        & $m_{\pi}$ & $m_{\bar{K}}$ & $m_{N}$ & $m_{\Sigma}$\\
\hline
        QCDSF-UKQCD & 203.7* & 482.0 & 965.7 & 1162.8 \\
      CLS D200 & 203.7 & 486.4 & 979.8 & 1193.9 \\
\hline\hline
    \end{tabular}
    \label{Masses of simulation results based on the QCDSF-UKQCD data and the recent lattice QCD data}
\end{table*}

\end{widetext}

\end{document}